\documentclass[a4paper]{jpconf}
\usepackage{graphicx}
\begin{document}
\title{ New X-ray views of the Galactic center observed with $Suzaku$}
\author{Katsuji Koyama, Yoshiaki Hyodo and Tatsuya Inui}
\address{ Department of Physics, Graduate school of Science, Kyoto University, 
Sakyo-ku, Kyoto 606-8502}
\ead{koyama@cr.scphys.kyoto-u.ac.jp}
\begin{abstract}

We report the diffuse X-ray emissions from the Sgr A and B regions
observed with $Suzaku$.  From the Sgr A region, we found many K-shell
transition lines of iron and nickel. The brightest are K$\alpha$ lines
from FeI, FeXXV and FeXXVI at 6.4~keV, 6.7~keV and 6.9~keV.  In
addition, K$\alpha$ lines of NiI and NiXXVII, K$\beta$ of FeI, FeXXV
and FeXXVI, and K$\gamma$ of FeXXV and FeXXVI are detected for the
first time. The center energy of K$\alpha$ of FeXXV favors collisional
excitation as the origin for this line emission. The ionization
temperature determined from the flux ratio of K$\alpha$ of FeXXV and
FeXXVI is similar to the electron temperature determined from the flux
ratio of K$\alpha$ and K$\beta$ of FeXXV, which are in the range of
5$-$7 keV. Consequently, the Galactic Center diffuse X-rays (GCDX) are
consistent with emission from a plasma nearly in ionization
equilibrium.  The radio complex Sgr B region also exhibits K$\alpha$
lines of FeI, FeXXV and FeXXVI.  The 6.7~keV line (FeXXV) map exhibits
a local excess at $(l,~~b) = (0^\circ.612,~~ 0^\circ.01)$, and could be a
new young SNR. The 6.4~keV image is clumpy with local excesses near
Sgr B2 and at $(l,~~b) = (0^\circ.74, ~~-0^\circ.09)$. Like Sgr
B2, this latter excess may be another X-ray reflection
Nebulae (XRN).
\end{abstract}

\section{Introduction}
The $Astro E2$ satellite was launched on 10th July in 2005 and was named
$Suzaku$ \cite{1} after an imaginary red bird in the ancient oriental
mythology.  The X-ray Imaging Spectrometer (XIS) \cite{2} is one of
the major instruments on $Suzaku$. The XIS consists of four sets of
X-ray CCD cameras, placed on the focal planes of the X-ray Telescope
(XRT) \cite{3}. Three XIS sensors contain front-illuminated (FI)
CCDs, and the other has a back-illuminated (BI) CCD \cite{2}.  The
other focal plane instrument is the micro-calorimeter (XRS) \cite{4},
which had extremely high energy resolution. The XRS was cooled down to
60 mK, and its overall energy resolution was originally 7 eV at 5.9 keV. 
These temperature and energy resolution were on-orbit world records. The
XRS was able to resolve, e. g., the iron He-like triplets, and hence should
have opened a new window of X-ray astronomy. However, regretfully, all
the liquid He evaporated very quickly.  We nevertheless still have the
XIS, which have the best performances among space CCD cameras. The
background level for diffuse sources, in particular at the energy of
5--10 keV range is nearly one order of magnitude lower than $Chandra$
and $XMM$. The effective area and energy resolution are also
comparable to or better than $XMM$. Therefore the XIS on $Suzaku$ is best
suited for observations of the diffuse X-rays from the Galactic Center
(GC). Early $Suzaku$ results will appear soon in the
refereed journal (\cite{5}, \cite{6}). This paper reports the summary
and essence of these papers.

\section{The Sgr A Region}

\begin{figure}[!ht] 
   \begin{center} \includegraphics[width=24pc]{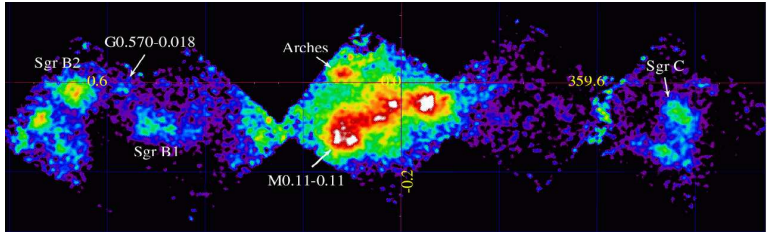} \caption{The
   narrow band map at the 6.4 keV line (the 6.34--6.46~keV
   band). Coordinates are galactic $l$ and $b$ in degrees.}
   \label{fig:6.4 keV}
\end{center}
\end{figure}

\begin{figure}[!ht] 
   \begin{center} \includegraphics[width=24pc]{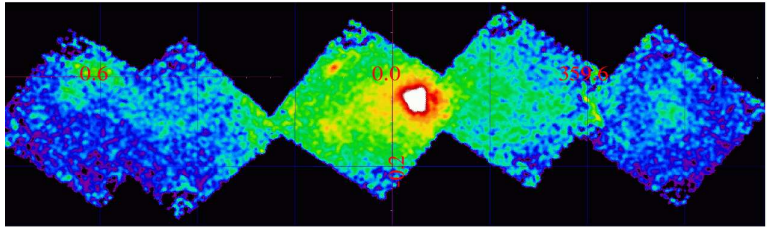} \caption{Same
   as figure \ref{fig:6.4 keV}, but for the 6.7 keV line (the
   6.62--6.74~keV band).}  \label{fig:6.7 keV}
\end{center}
\end{figure}

We have made narrow band GC maps in the 6.34--6.46~keV and
6.62--6.74~keV bands, which are shown in figures \ref{fig:6.4 keV} and
\ref{fig:6.7 keV}.  Although these line energies are very close to
each other, the X-ray maps are entirely different. The 6.4 keV line is
the K$\alpha$ of FeI, hence figure \ref{fig:6.4 keV} traces 
molecular clouds.  While the 6.7 keV line is K$\alpha$ of FeXXV, hence
figure \ref{fig:6.7 keV} traces very hot plasma.

\begin{figure}[!ht] 
   \begin{center} \includegraphics[width=18pc]{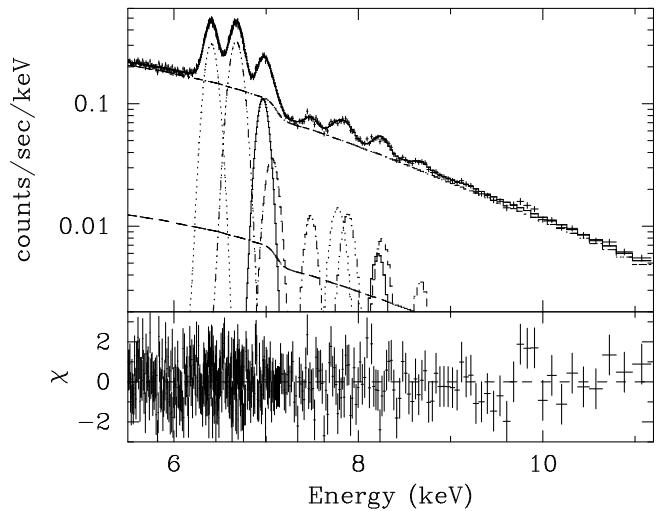}
   \caption{The averaged spectrum of the 3 FIs in the 5.5--11.5~keV
   band.  The data are taken from the full FOV excluding the
   calibration source regions and Sgr A East. The best-fit result is
   shown with the dot-dashed line. The long-dashed line is the model
   of the cosmic X-ray background (CXB) (after Koyama et al. 2006b,
   $PASJ$, submitted).}  \label{fig:hardspec}
\end{center}
\end{figure}

 \begin{table}[!ht] \begin{center} \caption{The detected lines from
 the Sgr A region. } \label{tabl:lines} \begin{tabular}{lc} \hline
 \multicolumn{2}{c}{Emission lines}\\ \hline Observed Center
 Energy$^*$ & Identification\\ (eV) & Line \\ \hline 6409 &
 FeI~K$\alpha$ \\ 6680 & FeXXV~K$\alpha$ \\ 6969 & FeXXVI~Ly$\alpha$
 \\ 7069 & FeI~K$\beta$ \\ 7490 & NiI~K$\alpha$ \\ 7781 &
 NiXXVII~K$\alpha$ \\ 7891 & FeXXV~K$\beta$ \\ 8220 & FeXXVI~Ly$\beta$
 \\ 8264 & FeXXV~K$\gamma$ \\ 8681 & FeXXVI~Ly$\gamma$ \\ \hline
 \end{tabular}\\ $*$ systematic line error is $^{+2}_{-7}$ eV.

\end{center}
 \end{table}

\subsection{The center energy and width of K$\alpha$ line of FeXXV}

The spectrum of the GCDX from the three FIs is given in figure
\ref{fig:hardspec}, where the non X-ray background (NXBG) made from
the night earth data is subtracted.  We fit the spectrum with a
phenomenological model (thermal bremsstrahlung + Gaussians +
absorption edge). The best-fit line center energies and the best line
candidates are given in table 1, where the possible systematic errors
of the line center energies are $^{+2}_{-7}$ eV.  The center energy of
K$\alpha$ of FeXXV is 6680$\pm1$ eV.  This line is a blend of
resonance, inter-combination and forbidden lines, hence the energy
depends on the flux ratio of these 3 lines.

For the origin of K$\alpha$ lines of FeXXV and FeXXVI, 
we discuss two possible cases: (1) collisional ionization and (2) recombining plasma,
while the latter case (case 2) includes a photo-ionized plasma and charge exchange processes.
The most distinct difference between the two cases is the flux ratio
of resonance and forbidden lines, such that collisional ionizing
plasma (1) emits the stronger resonance line than the forbidden line by
factor of 2, and vice versa for the recombining plasma (2). 
In fact,
the $Chandra$ and $XMM-Newton$ grating data of many stellar coronae (case 1)
have confirmed the relative dominance of the resonance line in
comparison to the forbidden line.  On the other hand, the case 2, 
the recombining of a low temperature electron is seen in a photo-ionized plasma, with the
typical case being NGC 1068 \cite{7}, which shows a larger flux of the
forbidden line than that of the resonance line.  Thus the center
energy of K$\alpha$ of FeXXV should be different for the two cases.

The center energy of the collisional ionization plasma (1) and that of the
charge exchange process (2) were measured from the iron ion beam
experiment in the laboratory \cite{8}).  The center energy depends on
the energy of iron beam and target materials.  Within a reasonable
range of conditions, the collisional ionization plasma (1) and the charge exchange process (2)
give center energies of 6680$-$6685 eV and 6666 eV, respectively.  
We also simulated the collisional-ionization case using
the plasma codes APEC and MEKAL, and also find central energies of
6680 eV (MEKAL) and 6685 eV (APEC), respectively.  Our observed energy
of 6680~eV agrees well with collisional ionization plasma of the
laboratory data and the MEKAL model, or at most, differ only 5~eV from
the APEC model.  The line width (1 $\sigma$) of K$\alpha$ of FeXXV is
determined to be 38~eV, exceeding the systematic effect of $\sim$30 eV
(the width of the calibration line). This large line broadening would
be due mainly to the triplet lines and satellite lines.

\subsection{Constraints on the cosmic ray iron velocity in the charge exchange scenario}

The other constraint on the charge exchange process is the Lyman limit
hump.  If an electron is captured to a level of a large principal
quantum number of an iron nucleus, the subsequent decay to the ground
state produces a hump at Lyman series limit ($\sim$ 9.2~keV).  However
we see no significant hump with the upper limit of $9\times
10^{-6}$~photons~s$^{-1}$~cm$^{-2}$ (we assumed a Gaussian line with
width 30~eV), which is about 6\% of the intensity of Ly$\alpha$.  This
upper limit sets the lower limit of the collision energy to $>$ 100~eV
amu$^{-1}$ \cite{9}, \cite{10}.  Thus the velocity of bare iron nuclei
must be larger than $\sim$150~km~s$^{-1}$, or the line width of
K$\alpha$ of FeXXVI (Ly$\alpha$) must be larger than 3~eV.  Since the
observed line width is smaller than that of the calibration line, the real line broadening should be
nearly zero. This is inconsistent with no hump at 9.2 keV.  We however
note that the accurate measurement of the line broadening of this
level ($\sim$ 3 eV) is very difficult with the XIS, hence further
confirmation with better resolution instruments is very important.

Together with the discussion of the previous section, we conclude that
the 6.7~keV and 6.9~keV lines in the GCDX are caused by collisional
ionization and do not originate from the charge exchange process.

\subsection{The flux ratio of  K-shell transition lines of FeXXV, XXVI and NiXXVII}

The ionization temperature determined from the best-fit K$\alpha$ flux
ratio between FeXXVI and FeXXV is similar to the electron temperature
determined from the best-fit flux ratio of K$\beta$ and K$\alpha$ of
FeXXV, within the error range of 5--7~keV.  The observed flux ratios of
K$\beta$ and K$\alpha$ of FeXXVI, and K$\gamma$ and K$\alpha$ of
FeXXV, XXVI are all consistent with an electron temperature of
$\sim5-7$~keV.  With the assumption that the relative abundances of
iron and nickel are proportional to the solar value, the K$\alpha$
ratio between NiXXVII and FeXXV is consistent with an the ionization
temperature of $\sim$5~keV.  We therefore conclude that the
temperature of the GC is $\sim 5-7$ keV in collisional ionization
equilibrium.

\section{The Sgr B Region}

\begin{figure} 
  \begin{center} \includegraphics[width=18pc]{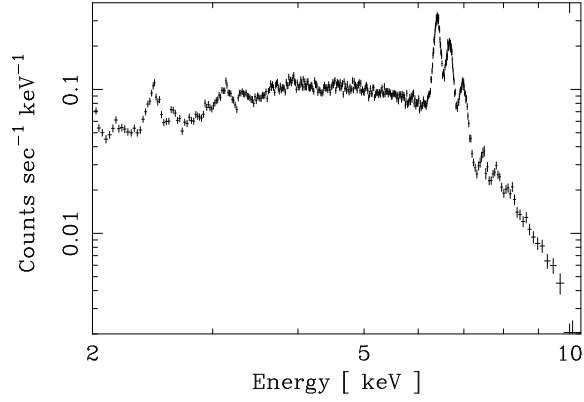}
  \end{center} \caption{The X-ray spectra from the full FOV of the
  XIS, but the CCD corners irradiated by the build-in calibration
  sources are excluded. All the four XIS data are co-added}
  \label{fig:sgrb-overall}
\end{figure}

The X-ray spectrum of all the Sgr B region is given in figure
\ref{fig:sgrb-overall}. The spectra of the four XISs (XIS0-XIS3) are
co-added and the NXBG is subtracted. With the superior energy
resolution of the XIS for diffuse sources, we can clearly resolve the
6.4~keV, 6.7~keV and 6.9~keV lines.

\begin{figure}
  \begin{center} \includegraphics[width=18pc]{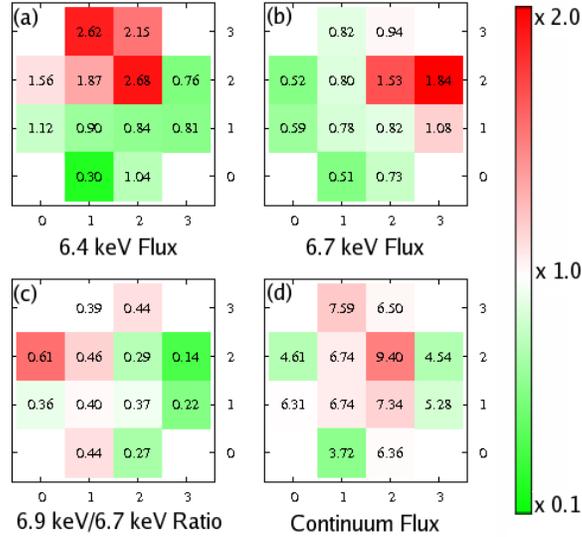}
  \end{center} \caption{The 6.4~keV (a), 6.7~keV (b), the flux ratio
  of 6.9~keV/6.7~keV (c), and the continuum flux map (d). The embedded
  numbers are fluxes in units of $10^{-5}$ photons cm$^{-2}$ s$^{-1}$
  arcmin$^{-2}$ or flux ratios. The color means that the average value
  is white, and the increasing (decreasing) excess above (below) the
  average is given with increasing deepness of red (green), where the
  maximum (minimum) is 2.0 (0.1) times of the average value.}
\label{fig:sgrb-colormap}
\end{figure}

In order to study this region more quantitatively, we have divided the XIS FOV
into 16 ($4 \times 4$) areas as is shown in figure
\ref{fig:sgrb-colormap} by the solid grids. For brevity, we assign
coordinates for each square position such as (0, 0) and (3, 3), where
the former and latter are lower-left and upper-right positions,
respectively. Since the satellite roll-angle of the observation was
nearly 0$^\circ$, the north of this coordinate is also the north of
the sky coordinate. The four corners, (0, 0), (0, 3), (3, 0) and (3,
3) contain the MnI K$\alpha$ and K$\beta$ line from the built-in
calibration sources, so we exclude these 4 areas in the
following analysis.

We then made X-ray spectra from the 12 squares. The NXBG spectra are
constructed from night earth data in the same detector area of the
12-square regions and subtracted from the spectra of the 12 Sgr B
regions.  The NXBG-subtracted spectra are fitted with a model of
thermal bremsstrahlung plus Fe absorption edge at the 7.1~keV and four
emission lines, which are the K$\alpha$ lines from FeI (6.4~keV),
FeXXV (6.7~keV) and FeXXVI (6.9~keV) and the K$\beta$ line from FeI
(7.06~keV).  Using the best-fit fluxes, we made color code maps
(figure \ref{fig:sgrb-colormap}), which are the fluxes of 6.4~keV
(figure \ref{fig:sgrb-colormap}a), 6.7~keV (\ref{fig:sgrb-colormap}b),
the flux ratio of 6.9~keV/6.7~keV (\ref{fig:sgrb-colormap}c), and the
continuum flux (\ref{fig:sgrb-colormap}d).

Both the 6.4 and 6.7~keV line maps show a large contrast in red and
green colors, which means that the fluxes have large variation from
position to position. On the other hand, the 6.9~keV/6.7~keV ratio is
rather smooth (the contrast of colors is weak) except the positions of
(3, 2) and (0, 2). If we exclude these positions, the mean value of
the other 10 squares is 0.36.  For a plasma in collisional ionization 
equilibrium (CIE), this value corresponds to a temperature of 6--7~keV.
Koyama et
al. (2006b) reported that the mean temperature of the Galactic Center
diffuse X-ray emission (GCDX) is 5--7 keV.  We thus conclude that the
origin of the diffuse emission is the same. At the Sgr B region, the
surface brightness of the GCDX is about a half of that in Sgr A.

The significant excess of the 6.7~keV flux and deficit of the
6.9~keV/6.7~keV flux ratio at the position of (3, 2) indicates that an
extra component with strong 6.7~keV line but no strong 6.9~keV line is
present at this position. The rather smooth distribution of the
continuum X-ray fluxes (see figure \ref{fig:sgrb-colormap}d) also
suggests that the excess emission is mainly due to the 6.7~keV
line. Since the 6.7 keV line is most likely the K$\alpha$ of FeXXV, a
probable origin of the local 6.7~keV excess is a hot plasma
with a temperature lower than the large scale GCDX. The
6.9~keV/6.7~keV flux ratio at the position (3, 2) is 0.14, which can
be converted to an ionization temperature of $\sim$4--5~keV.  This
temperature is typical for a young SNR like Sgr A East.  The 6.4~keV
flux (figure \ref{fig:sgrb-colormap}a) shows local excesses at the
position of Sgr B2 (2, 2) and the east of Sgr B2 (1, 3). The presence
of the 6.4~keV line is strong evidence for dense and cool molecular
clouds.

\subsection{Discovery of a new young SNR}

\begin{figure}
  \begin{center} \includegraphics[width=18pc]{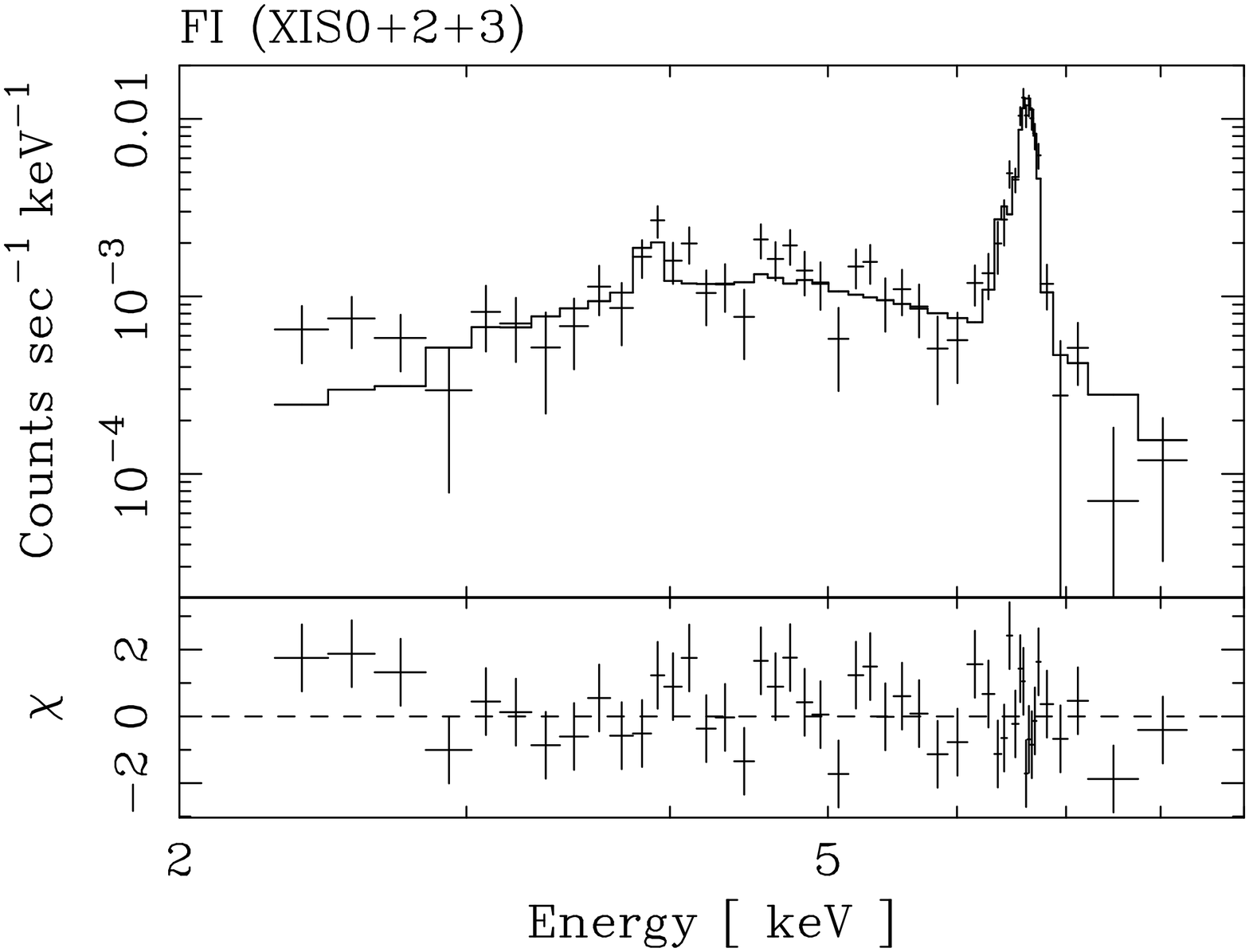}
  \end{center} \caption{The X-ray spectrum of the sum of 3 FI CCDs for
  a new SNR (G0.61+0.01) with the best-fit VNEI model (after Koyama et
  al. 2006c, $PASJ$, submitted).}
\label{fig:sgrb-g061001-spec}
\end{figure}

 Since the 6.7 keV excess at the position of (3, 2) is likely to be a new
SNR (see discussion below), we designate this source as Suzaku J1747.0-2824.5 (G0.61+0.01) 
from its center position.
The background-subtracted
 spectrum of G0.61+0.01 is given in figure \ref{fig:sgrb-g061001-spec}, for the 3
 FIs case, where the background was taken from the southeast region. 
We see pronounced peak at 6.7~keV, but the line shape is
 asymmetric with a tail at lower energy. This tail would be due to the
 presence of a faint K$\alpha$ line of FeI at 6.4~keV.

The FI and BI spectra are simultaneously fitted with a thin thermal
plasma in non-equilibrium ionization (VNEI)
\cite{11}, adding two Gaussian lines at 6.4~keV and 7.06~keV. 
These two lines represent the K$\alpha$ and K$\beta$ lines of FeI, 
where the flux of latter line is fixed to 12.5\% of the former. 
The best-fit result for the FI is  shown in figure \ref{fig:sgrb-g061001-spec}.
The best-fit temperature is $\sim$3~keV and the abundances of Ca and
Fe are 3-4 times of solar. The $N_{\rm H}$ value is $1.6\times
10^{23}$ cm$^{-2}$, larger than the typical value to the GC
($6\times 10^{22}$~H~cm$^{-2}$) \cite{12}.  Therefore, G0.61+0.01
would be located behind or in the rim of the Sgr B2 cloud. Since
G0.61+0.01 is located in the south of an expanding radio shell
\cite{13}, which is probably interacting with the Sgr B2 cloud rim, we
assume that the distance of G0.61+0.01 is the same as Sgr B2 and to be
8.5~kpc \cite{14}. Then the G0.61+0.01 size of about 2.4~arcmin 
(the radius of the major axis)
corresponds the real size of 6~pc. Dividing this size by the sound
velocity of the 3~keV plasma, we obtain the dynamical time scale of
$\sim 4\times10^{3}$ years. The 2--10 keV band luminosity is estimated
to be $1.5\times 10^{34}$ ergs s$^{-1}$. These values are typical for
an ejecta dominanted young SNR.

Since G0.61+0.01 is found at the edge of the XIS field, some parts of
this candidate SNR would be out of the XIS field. One possibility is
that G0.61+0.01 is a part of the expanding radio shell discovered by
Oka et al.\ \cite{13}. 
The kinetic energy of the radio shell is a few of $10^{52}$
erg s$^{-1}$, within the range of single or multiple supernova
explosions.

\section{Discovery of a New XRN}

\begin{figure}
  \begin{center} \includegraphics[width=18pc]{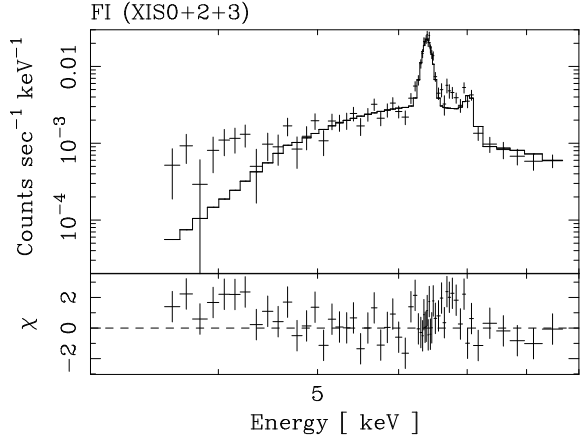}
  \end{center} \caption{The X-ray spectrum of the sum of the 3 FI CCDs
  for Sgr B2 with an absorbed power-law model and two Gaussian lines
  (after Koyama et al 2006c, $PASJ$, submitted).}
  \label{fig:sgrb-B2-spec}
\end{figure}

\begin{figure}
  \begin{center} \includegraphics[width=18pc]{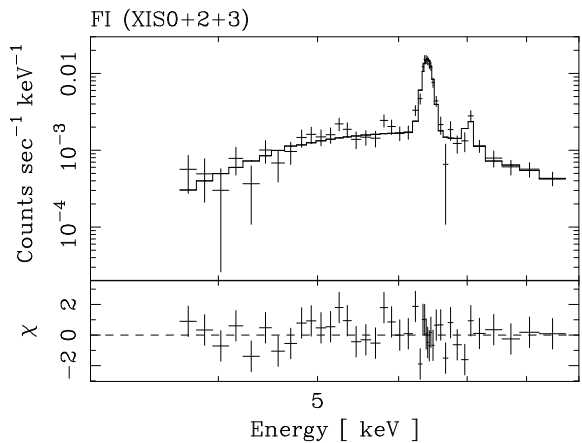}
  \end{center} \caption{Same as figure \ref{fig:sgrb-B2-spec}, but for
  a new source M0.74-0.09 (after Koyama et al. 2006c, $PASJ$,
  submitted).}  \label{fig:sgrb-m074009-spec}
\end{figure}

The 6.4~keV map shows two bright regions at (2, 2) and (1, 3). One is
Sgr B2, which has been already found as a strong 6.4 keV source
\cite{15}, and the other is a newly discovered source
Suzaku~J1747.7$-$2821.2 (M0.74$-$0.09). We show the background-subtracted
spectra of M0.74$-$0.09 and Sgr B2 in figures \ref{fig:sgrb-B2-spec}
and \ref{fig:sgrb-m074009-spec}. The latter is for comparison to the
former new source.  We simultaneously fit the FIs and BI spectra with
a model of absorbed power-law plus two Gaussians near at 6.4 and
7.06~keV, which are for the K$\alpha$ and K$\beta$ lines of FeI.  This
model nicely fits the data except an excess near the 6.7~keV line in
the Sgr B2 spectra.  One possibility is that the 6.7~keV enhancement
is a part of the new SNR candidate G0.61+0.01, because it is located
in the close vicinity of Sgr B2. The other possibility is that the
6.7~keV enhancement is due to high mass young stellar objects (YSO)
embedded in the center of Sgr B2, because some of the YSOs shows a
hint of the 6.7~keV line emission \cite{19}.

The Sgr B2 cloud has been studied extensively with $ASCA$ and
$Chandra$ \cite{15},\cite{17}. The authors concluded that the 6.4~keV
emission is due to a fluorescent by strong X-rays coming from Sgr
A$^{*}$, hence named the X-ray Reflection Nebula (XRN). $Suzaku$ found
a clear K$\beta$ line at 7.06~keV of the $\sim$10\% of K$\alpha$ and
deep Fe edge at 7.1~keV.  These discoveries provide additional
supports for the XRN scenario of Sgr B2.

The spectrum of M0.74-0.09 exhibits strong 6.4~keV line with an
equivalent width of 1.6~keV, 7.06~keV line and edge structure at
7.1~keV. All these lines and edge energies are consistent with being
from K$\alpha$, K$\beta$ and K$_{\rm edge}$ of FeI. The flux of the
7.06~keV line is about 10\% of that of the 6.4~keV line, which is also
consistent with the fluorescent X-rays. Unlike Sgr B2, no hint of YSOs
is found so far.  No bright point source is found in the $Chandra$
image. Therefore, the X-rays cannot be the scattering and fluorescence
by embedded YSOs or any internal source.  If the X-rays from Sgr B2
and M0.74$-$0.09 are due to the Thomson scattering and fluorescence of
the same irradiating external source like Sgr A$^*$, then the $N_{\rm
H}$ ratio between these sources should be same as the 6.4~keV line
flux ratio. The observed $N_{\rm H}$ ratio and the 6.4~keV line flux
is in good agreement with each other, hence supports the fluorescent
scenario. Therefore the XRN scenario successfully applied to Sgr B2
can be also applied to M0.74$-$0.09.

The counter scenario against the XRN is that the 6.4~keV line emission
is produced by the collision of electrons with neutral iron. Since the
cross section of iron K-shell ionization is maximum at the electron
energy of a few 10~keV \cite{18}, the most probable source is low
energy electrons (LEE) as proposed for the origin of the Galactic
Ridge iron K-shell emission \cite{19}. Since a few 10~keV electrons is
absorbed in less than $10^{22}$H cm$^{-2}$ of depth \cite{18}, the
produced X-ray spectrum has no large absorption edge. Our observation,
however, shows clear absorption of (4--10) $\times
10^{23}$H~cm$^{-2}$, in far excess to the Galactic interstellar
absorption \cite{12}. Thus the LEE origin is unlikely, unless we
assume a special geometry such that the 6.4keV source is deep in or
behind the dense cloud.

\ack
The authors express sincere thanks to all the $Suzaku$ team members
and our Laboratory colleagues. Y.H. and H.N. are supported by JSPS
Research Fellowship for Young Scientists.  This work is supported by
the Grant-in-Aid for the 21st Century COE "Center for Diversity and
Universality in Physics" from the Ministry of Education, Culture,
Sports, Science and Technology (MEXT) of Japan.

\end{document}